# An axiomatic method for studying the truth or falsity of the Hirano-Utsu law describing aftershocks

A.V. Guglielmi

*Schmidt Institute of Physics of the Earth, Russian Academy of Sciences; Bol'shaya Gruzinskaya str., 10, bld. 1, Moscow, 123242 Russia; guglielmi@mail.ru*

**Abstract**. The power law of aftershock evolution was proposed by Hirano in 1924 and introduced by Utsu into seismology in the second half of the last century. The Hirano-Utsu law is widely used in studying the relaxation of earthquake source after the main shock of an earthquake. The prevailing view in the literature is that the Hirano-Utsu law is an improved version of Omori's hyperbolic law, formulated in 1894. The author disagrees with this notion. The paper proposes an axiomatic approach to the study of aftershocks. A phenomenological parameter of the source, called the deactivation coefficient, was introduced. The theory is based on axioms that do not contain any a priori statements regarding the form of the law of aftershock evolution. Formulas for the deactivation coefficient are derived from the axioms, allowing one to experimentally establish the truth or falsity of the Hirano-Utsu and Omori laws. A two-stage mode of source relaxation was discovered. In the first stage, called the Omori epoch, the Omori law is strictly followed. The Omori epoch ends with a bifurcation, after which aftershock activity becomes unpredictable. Omori's law is not fulfilled at the second stage of evolution. The Hirano-Utsu law is not fulfilled either at the first or second stage.

## 1. Introduction

The spontaneous formation of a main discontinuity in rocks at the source manifests itself in the form of the main shock of an earthquake. The equilibrium of the source is sharply disrupted. After this, a long process of relaxation of the source begins to a new state of equilibrium. Relaxation is accompanied by secondary discontinuities in the source volume, which are recorded by seismographs in the form of aftershocks, the frequency of which $n(t)$ gradually decreases over time.

In 1924, Hirano [1] proposed formula

$$n(t) = \frac{k}{(c+t)^P} \qquad (1)$$

to describe the average frequency of aftershocks. Here $k>0$, $c>0$, $t\geq 0$, $p>0$. In 1938, Jeffreys [2] referred to formula (1), but, in general, it did not attract much attention until the second half of the last century, when Utsu [3–5] with his research demonstrated the convenience and effectiveness of formula (1) in approximating data observations. After this, formula (1) acquired the status of the law of aftershock evolution in the seismological literature. It is called the Omori-Utsu law [6], Utsu law [7], or modified Omori law [8, 9]. In my opinion, it would be correct to call formula (1) the Hirano-Utsu law.

The prevailing idea in the literature is that the Hirano-Utsu law is an improved version of Omori's law [10]

$$n(t) = \frac{k}{c+t}, \qquad (2)$$

formulated in 1894. This paper presents arguments in favor of the idea that formula (1) is not applicable to describe aftershocks, while formula (2) is applicable, but only at the first stage of source relaxation.

It would seem that this is impossible, since formula (1) contains an additional parameter $p$ for fitting and therefore certainly approximates the observational data better than formula (2). And in general, the judgment of a non-specialist (I am a radio physicist by training) about the fallacy of the well-known law of seismology (1) may seem not worth attention. But the point here is this.

Over the past few years, I have been working on the problem of aftershocks as part of a research team that includes seismologist A.D. Zavyalov and geophysicists B.I. Klain and O.D. Zotov. The result of productive work is reflected in review papers [11–15] (see also [16–23]). In particular, we were able to discover that none of the above formulas describes the evolution of aftershocks holistically.

Our approach to processing and analysis of aftershocks was based on the concept of the source as a dynamic system, the state of which is characterized by a phenomenological parameter $\sigma(t)$, which we called the source deactivation coefficient. The value of $\sigma(t)$ was determined as the solution to the inverse problem for the evolution equation [24, 25]

$$\frac{dn}{dt}+\sigma n^2 = 0. \qquad (3)$$

Having accepted (3) as a postulate, we obtained a number of interesting and previously unknown results. However, our radical conclusion about the inapplicability of formula (1) for describing aftershocks may still raise doubts for the following reason. Namely, for the simple reason that equation (3) does not have the property of an axiom, i.e. is not an obvious statement that does not require proof. Therefore, in this paper we will base on axioms that do not contain any a priori statements like formulas (1), (2) or equation (3). As theorems, we will obtain formulas for the deactivation coefficient, allowing us to experimentally establish the truth or falsity of the Omori and Hirano-Utsu laws.

## 2. Axmomatic justification of Omori's law

Since formula (2) is simpler than (1), we first axiomatize Omori's theory. The theory is based on two axioms:

1. The deactivation coefficient can be calculated from aftershock frequency measurements.

2. The deactivation coefficient is independent of time if and only if the aftershock frequency obeys Omori's law.

The theorem clearly follows from the axioms:

$$\sigma(t) = -\frac{1}{n^2(t)} \frac{dn(t)}{dt}. \tag{4}$$

It would seem that the problem of calculating the deactivation coefficient from observational data on the frequency of aftershocks has been solved, but this is not entirely true. In physical and mathematical sciences, it happens that a fundamental possibility is associated with practical impracticability, or with the inexpediency of realizing the opportunity due to the complete unproductivity of the result. We encountered exactly such a case.

We need one more theorem: the evolution of aftershocks is described by the linear differential equation

$$\frac{d}{dt} g(t) - \sigma(t) = 0. \tag{5}$$

Here $g(t) = 1/n(t)$ is an auxiliary function. Without stopping at the trivial proof of the theorem, we pose the inverse problem for the evolution equation: calculate the deactivation coefficient from the measurement data of the aftershock frequency. In this form, the problem is formulated incorrectly, as often happens in geophysics when inverse problems are formulated. In our case, regularization consists of adequate smoothing of the auxiliary function. As a result, the solution takes on the form

$$\sigma(t) = \frac{d}{dt}\langle g(t) \rangle, \qquad (6)$$

where the angle brackets indicate the smoothing procedure.

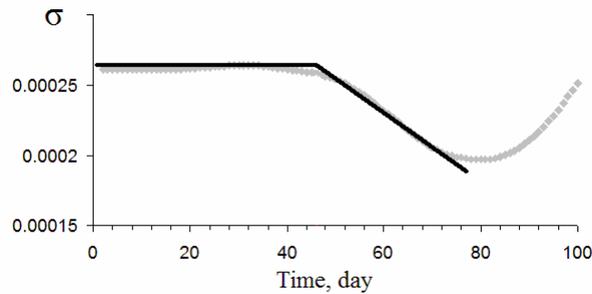

Illustration of the two-stage mode of aftershock evolution. The main shock with magnitude $M = 7.1$ occurred in Southern California on October 16, 1999. The smooth line shows the variation of the deactivation coefficient. Straight line segments show the piecewise linear approximation of the Omori epoch and the beginning of the second stage of evolution.

The method of processing aftershocks using formula (6), axiomatized above, made it possible to detect a two-stage evolution regime [12–15, 20]. At the first stage $\sigma = \text{const}$, i.e. the classical Omori law is strictly followed (see figure). In honor of Fusakichi Omori, the first stage was called the *Omori epoch*. The duration of the Omori epoch varies from case to case from approximately 10 to 100 days. The deactivation coefficient in the Omori epoch is lower, the greater the magnitude of the main shock.

The Omori epoch ends with bifurcation, after which the second stage of the source evolution begins. At the second stage, Omori's law does not work. The transition from the first stage of evolution to the second is quite abrupt, as can be seen in the figure.

Let us introduce the parameter $\theta = d\sigma/dt$, which characterizes the rate of change of the deactivation coefficient over time. During the Omori epoch, $\theta = 0$. At the moment of bifurcation, the parameter experiences a jump by a finite value $\Delta\theta$. The value $\Delta\theta$ can be either positive or negative.

Thus, analysis of field data using our method gives the researcher three new parameters: the deactivation coefficient $\sigma$ in the Omori epoch, the duration of the Omori epoch, and the jump $\Delta\theta$ at the moment of bifurcation. This enriches the tools suitable for searching for patterns of source evolution.

After bifurcation, the deactivation coefficient begins to change over time in the most unpredictable way. This vaguely resembles the transition from laminar to turbulent fluid flow, known in hydrodynamics. We do not know the law of source dynamics at the second stage of evolution.

### 3. Inapplicability of the Hirano-Utsu formula

Let us now axiomatize the Hirano-Utsu theory. The deactivation coefficient of the source will be denoted by the symbol $\sigma'$. We use the following system of axioms:

1. The deactivation coefficient can be calculated from aftershock frequency measurements.

2. The deactivation coefficient is independent of time if and only if the aftershock frequency obeys the Hirano-Utsu law.

In other words, the first axiom is left unchanged, and the second axiom is changed in such a way that the property $\sigma' = \text{const}$ serves as a criterion for the validity of the Hirano-Utsu law.

Similar to how it was done above, from the new system of axioms we derive a formula for calculating the deactivation coefficient:

$$\sigma'(t) = \frac{1}{p} \langle g(t) \rangle^{\frac{1}{p}-1} \frac{d}{dt} \langle g(t) \rangle. \tag{7}$$

When $p=1$, formula (7) coincides with formula (6). Applying formula (7) at $p=1$ to the analysis of aftershocks, we will come to the conclusion that there are two stages in the evolution of aftershocks. At the first stage, Omori's law is strictly followed, and at the second stage, the frequency of aftershocks changes over time in an unpredictable way.

Let us show that when $p \neq 1$ the Hirano-Utsu law does not apply either at the first or at the second stage of evolution. At the first stage $d\langle g \rangle / dt = \text{const}$, and from formula (7) it follows that

$$\sigma'(t) = \text{const} \cdot t^{\frac{1}{p}-1}. \tag{8}$$

When $p<1$, the deactivation coefficient monotonically increases, and when $p>1$, it monotonically decreases over time. Thus, at the first stage, the Hirano-Utsu law is not satisfied. The law is also not fulfilled at the second stage of evolution. In fact, both quantities $\sigma$ and $\sigma'$ after bifurcation change unpredictably over time. Therefore, after bifurcation, neither Omori's law nor the Hirano-Utsu law work.

## 4. Discussion

It is necessary to explain our choice of basic postulates. The theory of aftershocks, like any physical theory, is a phenomenological theory. This means that any phenomenological parameter that appears in the theory must either be derived within the framework of a higher-level phenomenological theory, or a method for calculating it based on measurements must be indicated. The following historical example will illustrate the point. In thermodynamics, for a long time they could only measure the temperature of a body, until Maxwell indicated the derivation of temperature within the framework of the kinetic theory. We cannot derive the deactivation coefficient from a higher level phenomenological theory of aftershocks. Such a theory has not yet been created. We can only obtain information about the value of the deactivation coefficient based on measurements. Thus, the first axiom specifies the general requirement for the phenomenological parameter of a physical theory.

The choice of the second axiom is motivated by the research problem. It's quite understandable. If it turns out that over any finite period of time the deactivation coefficient is variable, then this will mean that Omori's law (or the Hirano-Utsu law) does not work. The criterion for the applicability of Omori's law (Hirano-Utsu) is the condition $\sigma = \text{const}$ ($\sigma' = \text{const}$) at a given specific time interval during the evolution of aftershocks. Taken together, two pairs of axioms form the evidentiary basis for verification of the Omori and Hirano-Utsu laws.

Let us note that in our notation the evolution of aftershocks is described by differential equation (5) of the simplest form. It is easy to verify that linear equation (5) is equivalent to the simplest nonlinear equation (3). Both equations allow natural generalizations, one of which was successfully used in searching for patterns in the spatiotemporal evolution of aftershocks [26, 27].

### 5. Conclusions

1. The earthquake source is presented as a dynamic system, the state of which is characterized by the deactivation coefficient.

2. The axiomatized method for processing and analyzing aftershocks is proposed.

3. The inapplicability of the Hirano-Utsu formula for describing aftershocks has been proven.

*Acknowledgments*. I express my gratitude to B.I. Klain, A.D. Zavyalov and O.D. Zotov for many years of cooperation in the study of aftershocks. I thank A.S. Potapov and F.Z. Feygin for their attention and support.